**One-Dimensional Materials Supported in Two-Dimensional van der Waals Metal-Organic Frameworks with Optical Anisotropy Switching via Twist-Engineering**

*Eleni C. Mazarakioti,† Carla Boix-Constant,† Iván Gómez-Muñoz,† Diego López-Alcalá, Sergio Revuelta, Marco Ballabio, Vasileios Balos, José J. Baldoví, Enrique Cánovas, Josep Canet-Ferrer, Guillermo Mínguez Espallargas, Samuel Mañas-Valero,* Eugenio Coronado**

E. C. Mazarakioti, C. Boix-Constant, I. Gómez-Múñoz, D. López-Alcalá, J. J. Baldoví, J. Canet-Ferrer, G. Mínguez Espallargas, S. Mañas-Valero, E. Coronado
Instituto de Ciencia Molecular (ICMol), Universitat de València, Catedrático José Beltrán 2, Paterna, 46980 Spain.
† E. C. Mazarakioti, C. Boix-Constant and I. Gómez-Múñoz contributed equally to this work.
E-mail: samuel.manas@uv.es, eugenio.coronado@uv.es

S. Revuelta, M. Ballabio, V. Balos, E. Cánovas
IMDEA Nanociencia, Campus Universitario de Cantoblanco, Faraday 9, 28049 Madrid, Spain.

## Abstract

Van der Waals (vdW) materials provide a platform to study and control the physical properties of low-dimensional materials. While strategies developed for two-dimensional (2D) crystals are not directly transferable to one-dimensional (1D) systems, we can benefit from them by creating layers formed by interconnected chains. Here, we develop a molecular strategy to illustrate this concept consisting of assembling 1D materials in 2D metal-organic frameworks (MOFs). Crystals of [FeX(pzX)(bpy)] (X = Cl, F; pz = pyrazole; bpy = bipyridine) consist of iron chains along the *b*-axis, crosslinked *via* bpy ligands along the *a*-axis to form 2D layers, stacked along the *c*-axis *via* vdW forces. This structural anisotropy manifests itself in highly-anisotropic optical properties, as demonstrated by optical measurements in the visible and terahertz ranges, results which are supported by DFT calculations. Chemical substitution enables the tuning of the optical properties, as exemplified by the photoluminescence of the Cl-derivative, which is quenched for the F-derivative. Thin-layers are obtained by mechanical exfoliation, and their optical properties are further tuned through the fabrication of orthogonally-twisted vdW heterostructures, enabling to effectively switch-off the optical anisotropy. Our work highlights the chemical flexibility of vdW layered MOFs as a platform for designing and manipulating 1D architectures.

## 1. Introduction

Two-dimensional (2D) materials with weak van der Waals (vdW) interactions between the layers (*aka.* vdW materials) are an important topic of interest in materials science, unmatched in terms of the possibilities they offer for their manipulation using both chemical and physical approaches.[1–7] Thus, ultrathin layers of these materials (down to the single-layer limit) can be easily obtained *via* exfoliation or direct growth techniques,[8] which can be stacked and twisted at will generating an infinite number of vdW heterostructures with novel and tunable properties.[9–13] An extension of this concept towards lower dimensionalities remains largely unexplored.[14,15] Motivated by the recent physical developments observed in graphene nanoribbons,[16–19] the extension to one-dimensional (1D) systems is particularly appealing, as dramatic changes in properties arise in the 1D limit.[20–22] However, this also presents significant challenges, since the manipulation and control of 1D materials – both to study their distinctive properties and to integrate them with other materials to exploit these properties in devices – remains an open problem.[18,23–25] Some material candidates to overcome these challenges have been recently proposed *via* the use of 2D vdW materials of inorganic nature as platform to host 1D chains.[26,27] Still, this inorganic approach is limited by the difficulty in designing layers containing well-separated 1D chains. In this work, we propose a molecular approach that takes advantage of the versatility of coordination chemistry to design, by a proper selection of the ligands, vdW layered metal-organic frameworks (MOFs) containing well-isolated 1D metal-based chains.[28–32] Thus, we show that by reacting 4-Xpyrazolate ($pzX^-$; X = F and Cl) and 4,4'-bipyiridine (bpy) with $Fe^{2+}$, a layered structure formed by $Fe^{2+}$ chains cross-linked by bpy ligands is obtained. These neutral layers stack along the out-of-plane direction by vdW forces. Electronically, these MOFs are highly-anisotropic quasi-1D semiconductors, as probed by photoluminescence and optical conductivity measurements and corroborated by band structure calculations. In addition, using the well-stablished methods developed in the field of 2D materials to obtain and manipulate atomically-thin layers (micromechanical exfoliation and deterministic transfer), twisted heterostructures with tunable optical anisotropy are fabricated.

## 2. Results and Discussion

### 2.1. Chemical design and structural characterization of the materials

The 2D-supported MOF approach introduced here for manipulating the properties of 1D materials relies on two main requirements: (1) the assembly of molecular moieties in a manner that vdW neutral layers are formed, consisting of neutral chains of metal coordination complexes cross-linked by a neutral bridging ligand acting as spacer; and (2) the growth of layered MOF crystals of sufficient size and quality to allow exfoliation, handling, and characterization using techniques developed for 2D materials.

In coordination chemistry, the former requirement is difficult to reach since these chain structures are very often obtained as crystalline salts composed by anionic or cationic chains and charge compensating counterions, while the latter is limited by the preparation procedure.[32] This typically involves the crystallization of the coordination polymers in a solution containing the molecular precursors, resulting in the formation of poorly crystalline and insoluble powders with small particle size, which are impossible or extremely difficult to exfoliate.[33,34]

Despite the former difficulty, the synthesis of layered MOFs containing 1D chains has been previously exploited *via* different hydrothermal procedures.[35,36] A relevant example is provided by the series $MX_2$(bpy), formed by neutral layers containing chains of divalent metal centers ($M^{2+}$ = Fe, Co, Ni) octahedrally coordinated to four bridging halogen anions ($X^-$ = Cl, Br) and two bpy ligands at *trans* positions.[37–41] However, this approach does not allow to obtain crystals of suitable size and quality to be exfoliated and manipulated. In previous works, we showed that this limiting factor can be overcome using a solvent-free approach involving the appropriate sublimable molecular precursors (metallocenes and benzimidazole-type ligands) capable of releasing the metal ions in a controlled manner. This provided the first examples of layered magnetic MOFs able to be exfoliated down to the monolayer and magnetically characterized using micromechanical techniques.[42–44] Here, we initially followed this approach for growing crystals of the series $FeX_2$(bpy). Note that, in the present case, this approach is more complex since it also requires to select, besides the iron source (ferrocene), a halogen source. We have chosen 4-halopyrazoles (X= F, Cl, Br, and I) due to their ability to thermally decompose releasing the halogen, X.

The solvent-free reaction of -halopyrazole, ferrocene, and 4,4'-bipyridine in a sealed tube resulted (upon a thermal treatment) in the formation of two different families of compounds: with the heavier halogens (X= Br and I derivatives), this procedure affords crystals of the previously reported series $FeX_2$(bpy).[38,39] Still, no significant improvement in terms of size and quality has been observed as compared with the hydrothermal method. Hence, no further physical characterization and manipulation of these layered materials has been attempted. In turn, for the lighter halogens (X= Cl and F derivatives), a novel phase of brownish-transparent crystals of larger sizes and rectangular shapes (up to 3 mm for the long side of the rectangle in the Cl derivative, and up to 1 mm for the F derivative) are formed. Interestingly, in this phase the 4-halopyrazole is incorporated in the layered structure, together with the halide anion originated from the dissociation of the same molecule, as shown by single-crystal X-ray diffraction experiments. Both compounds crystallize in the monoclinic space group number 14 ($P2_1/n$ for the Cl-derivative and $P2_1/c$ for the F-derivative) and consist of the asymmetric unit [FeX(pzX)(bpy)]. These systems exhibit a layered structure where each layer comprises $Fe^{2+}$ centers coordinated by a combination of the halide anion ($X^-$), the anionic 4-halopyrazolate ligand ($pzX^-$) and the neutral 4,4'-bipyridine ligand (bpy). Specifically, the compounds comprise $Fe^{2+}$ ions exhibiting a slightly distorted octahedral geometry, each one coordinated by four nitrogen atoms originated from two pzX and two bpy ligands, as well as two halogen bridges forming $Fe^{2+}$ chains running along the *b* axis that are interconnected through bridging bpy ligands to afford a 2D layer material extending in the *ab* plane. The $Fe^{2+}$ centers are arranged in a grid-like pattern within each layer, with significant Fe–Fe distances ranging from 3.94 Å (in-chain) to 11.59 Å (inter-chain), reflecting the large spacing between metal centers in adjacent chain units. The layers are weakly interacting with vdW interactions between the halogen atom of the pyrazolate ligand and the halogen bridge, originating an interpenetrating layered structure (**Figure 1**). The structural details are reported in the **Supplementary Section 1**, together with the crystallographic data.

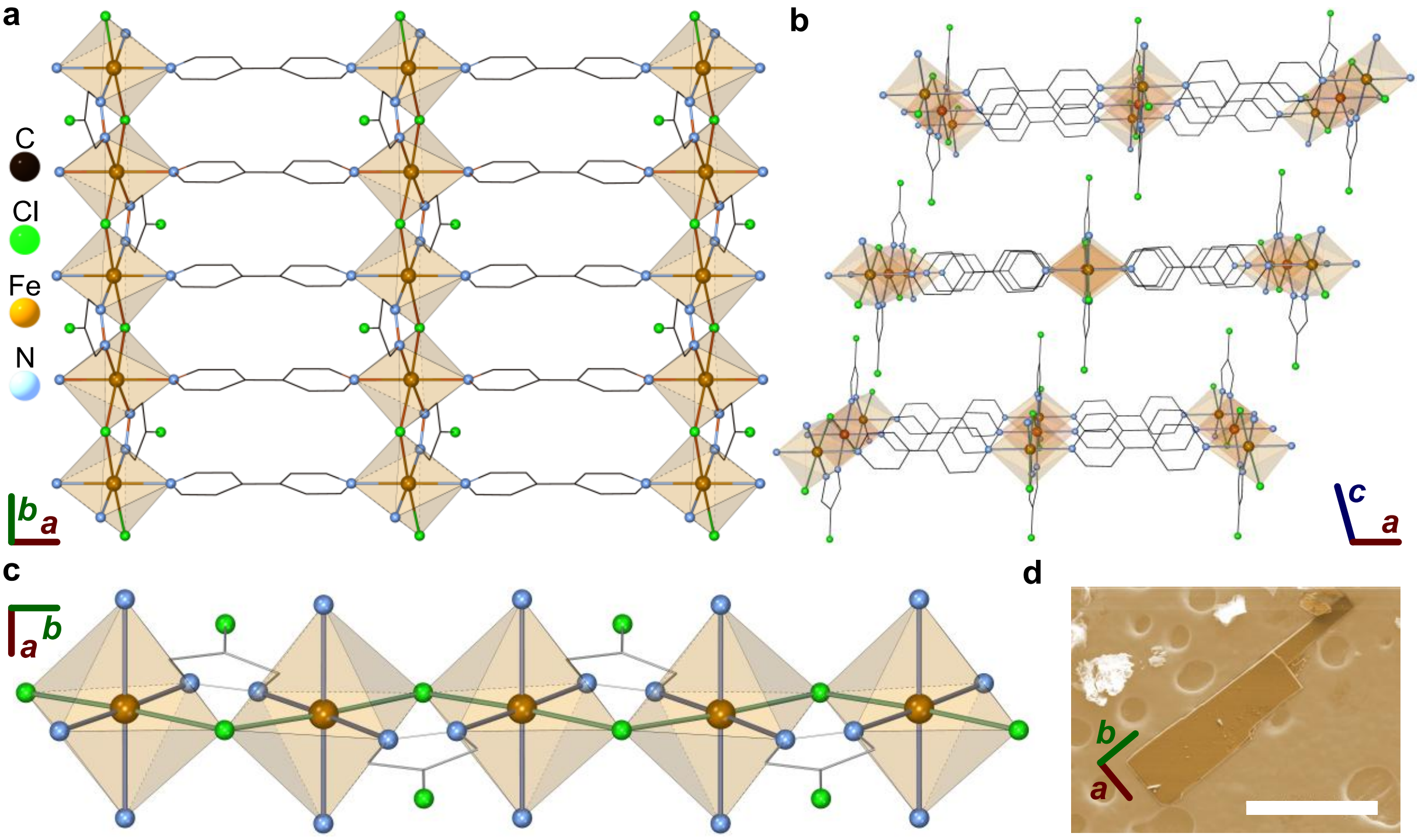

**Figure 1. Layered crystal structure of [FeX(pzX)(bpy)] ($X^-$ = Cl and F); $pzX^-$ = 4-halopyrazolate; bpy = 4,4'-bipyiridine).** a) In-plane structure of the layers showing the presence of well-separated $Fe^{2+}$ chains connected by bpy ligands. The iron ions are represented at the center of octahedral sites, while green and blue balls are halide and nitrogen coordinating atoms, respectively. b) Layered structure of the material highlighting the stacking of the layers. c) Detailed view of the $Fe^{2+}$ chain, where the zig-zag structure along the *b* axis formed by iron centers can be clearly visualized. d) Scanning electron microscope image exhibiting the layered facet for a Cl-derivative crystal. Scale bar: 500 μm

A final comment to conclude this section is to speculate about the reason why the novel pyrazole-containing structure, [FeX(pzX)(bpy)], is not obtained with the heavier halogens (Br and I derivatives), which prefer to form the $MX_2$(bpy) structure without integrating the 4-halopyrazole anion. Our working hypothesis is that it can be likely attributed to the increase in size of the Br and I atoms, as compared with F and Cl. It is well established that C–X bond strength decreases as the size of the halogen (X) increase. As a result, C–X bond scission becomes increasingly favorable along the series F → Cl → Br → I.[45–47] This tendency aligns with the observed loss of the pyrazolate ligand in the Br- and I-derivatives but incorporation into the framework for the weaker C–X bond (F- and Cl- derivatives). From a structural perspective, owing to this increase, the angle Fe–X–Fe becomes narrower, near 90°, and the Fe–X bond length increases up to 2.65 Å. Under these conditions it is not possible to accommodate the N–N group of the pyrazole keeping constant the Fe–N bond distance at 2.16 Å and a X–Fe–N angle of 90º in the octahedral Fe center (see **Supplementary Section 1**).

## 2.2. Optical properties of [FeX(pzX)(bpy)] (X = Cl, F).

The chemical design of MOFs has demonstrated its effectiveness in modulating physical properties such as magnetic anisotropy and dimensionality,[35,48–51] as illustrated by the early magnetic studies of the $MCl_2$(bpy) family (M = Fe, Co, Ni) and by subsequent investigations of closely related structures.[52] However, when it comes to optical anisotropy, examples within MOFs remain comparatively scarce.[53] The most notable feature observed in [FeX(pzX)(bpy)] (X = F, Cl) crystals is their marked optical anisotropy, showing clear distinct properties if measured along the metal chain direction (*b* axis) or perpendicular to it (*a* axis), as evidenced by photoluminescence (PL) measurements and by the high optical birefringence in the visible and THz range (**Figure 2**). To quantitatively measure this anisotropy, high-quality single-crystals with several millimeters of lateral size are required. Taking this into account, we focus primarily on the Cl-derivative, which has afforded much larger crystals to perform such measurements. When appropriate, the results are compared with those of the F-derivative.

In **Figure 2.a**, we show the PL spectra emitted by a single-crystal of the Cl-derivative, when the linear polarization of the incident excitation light source is oriented along *a* (red) or *b* (green) axes (as defined in **Figure 1**). The PL characterization is carried out using a laser operating at 532 nm wavelength (2.33 eV). The PL emission shows a broad feature centered around 1.96 eV, which is clearly more intense when the light polarization is aligned along *a*. Upon rotating the sample while keeping a fixed linearly polarized incident light (**Figure 2.b**), we can retrieve a typical behavior for a linearly polarized emitter: maximum (minimum) intensity along the *a* (*b*) axis (corresponding to 90 and 0 degrees, respectively), as seen in the polarization diagram of (**Figure 2.c**).[54] We can resolve as well prominent peaks in the high energy region of the spectra that correspond to the Raman scattering of vibrational modes. Raman and PL spectra upon different wavelengths are shown in the **Supplementary Section 2.1.**

To further investigate the optical anisotropy, we perform reflectivity measurements in the visible. We focus on a region above the PL emission to give a rough estimate of the refractive index of the material (**Figure 2.d**) in this range. Overall, the reflectivity of the bulk crystals is around 10% in the range of interest. In the case of thick crystals, the reflected light is higher when the incident excitation light is polarized along *a*. However, there is a range between 610-750 nm wavelength where the reflectivity is clearly higher for the light polarized along *b*. The combination of these data with optical transmissivity drives to an estimation of an average refractive index of n ~ 2 with variations around 10% depending on the polarization (details are shown in **Supplementary Section 2.2**). Interestingly, the system exhibits birefringence in the analyzed frequency window ($\Delta n$ in **Figure 2.d**, defined as $\Delta n = n_b - n_a$, being $n_a$ and $n_b$ the refractive index obtained for *a* and *b* directions). We obtain a maximum birefringence value of 0.3 at 507 nm. This is a value comparable with other vdW semiconductors in the visible range (as black phosphorous, $TiS_3$ or $As_2S_3$) or minerals like rutile or hematite, and higher than other birefringent crystals widely used for commercial purposes in optical components or laser devices, such as $YVO_4$ (a comparative table is given in the **Supplementary Section 2.3**).[55] This optical behavior is typical of quasi-1D systems as, for example, semiconductor nanostructures under quantum confinement or uniaxial strain,[56] which exhibit strong polarization dependence of the absorption and PL,[57] or other van der Waals materials such as CrSBr, $ReSe_2$, $Ta_2NiSe_5$ or black phosphorous, among others.[58–62]

We have also analyzed the optical response of the samples by THz time-domain spectroscopy. A crystal of [FeCl(pzCl)(bpy)] is placed in a rotary mount, and the transmittance of a freely propagating THz pulse (comprising photons from 0.5 to 3.0 THz) is measured at different crystal orientations (see **Experimental Section** and **Supplementary Section 3**). In line with the PL and visible analysis, we observe a large anisotropy in the response of the crystal at the probed THz frequencies. From the FFT analysis of the THz pulses, we clearly observe a change in the optical absorption at the THz frequencies (top panel in **Figure 2.e**). Specifically, three clear resonances at 1.2, 1.61 and 2 THz become visible along *a*. We tentatively attribute these features to molecular vibrations and/or rotations in the single crystal, which are selectively activated by the linearly polarized THz probe along *a* direction. In addition, we have retrieved the real part of the refractive index (middle panel in **Figure 2.e**; see **Supplementary Section 3** for details), from which we can resolve birefringence in the crystal at the probed frequencies (bottom panel in **Figure 2.e**). The maximum birefringence is 0.22 at 2.2 THz, a figure that is comparable with other birefringent materials (see table in **Supplementary Section 3.5**) and could be eventually exploited for THz retarders and waveplates in the field of THz photonics.[63–67] Finally, from the complex refractive index we extract the complex dielectric permittivity (**Supplementary Section 3**) and, from it, the frequency resolved complex conductivity of the sample (**Figure 3.f**). The conductivity also exhibits marked anisotropy when comparing the response along *a* (that is, along the bipyridine bridge) and *b* (along the Fe chains linked *via* chloride and Cl-pyrazolate anions). The DC conductivity at room-temperature is ~$10^{-8}$ S/cm, without significant anisotropy (see **Supplementary Section 5**). This conductivity value is comparable to those typically reported in MOFs (~$10^{-7}$–$10^{-10}$ S/cm)[68–70].

In contrast with the Cl derivative, the F derivative does not exhibit photoluminescence, independently of the orientation of the incident light polarization with respect to the crystallographic axes. Still, the pronounced anisotropy in the optical properties is qualitatively maintained, as shown by the reflectivity measurements in the visible range (**Supplementary Section 2.2**). Considering the response of a FeF(pzF)(bpy) single crystal at THz frequencies, a marked anisotropy on its optical response along its *a* and *b* crystal axis is observed as well. Comparing its response to the FeCl(pzCl)(bpy) crystal, marked differences are noted (**Supplementary Section 3.3**). Specifically, while the Cl derivative shows three clear resonances at 1.2, 1.61 and 2 THz along *a*, the F based crystal reveals a prominent single mode centered approximately at 1.5 THz that, in analogy with FeCl(pzCl)(bpy), can be associated with a soft vibrational mode.

In the broader context of MOFs, Khalil *et al.*[71] have recently reviewed different anisotropic MOF properties (we give a brief summary in **Table S7**). Still, to the best of our knowledge, optical anisotropy has been observed only in bulk MOF crystal based on copper(II) cations and a hybrid angle-shaped 1,2,4-triazole-carboxylate ligand with a rigid cage hydrocarbon (adamantane) linker.[72]

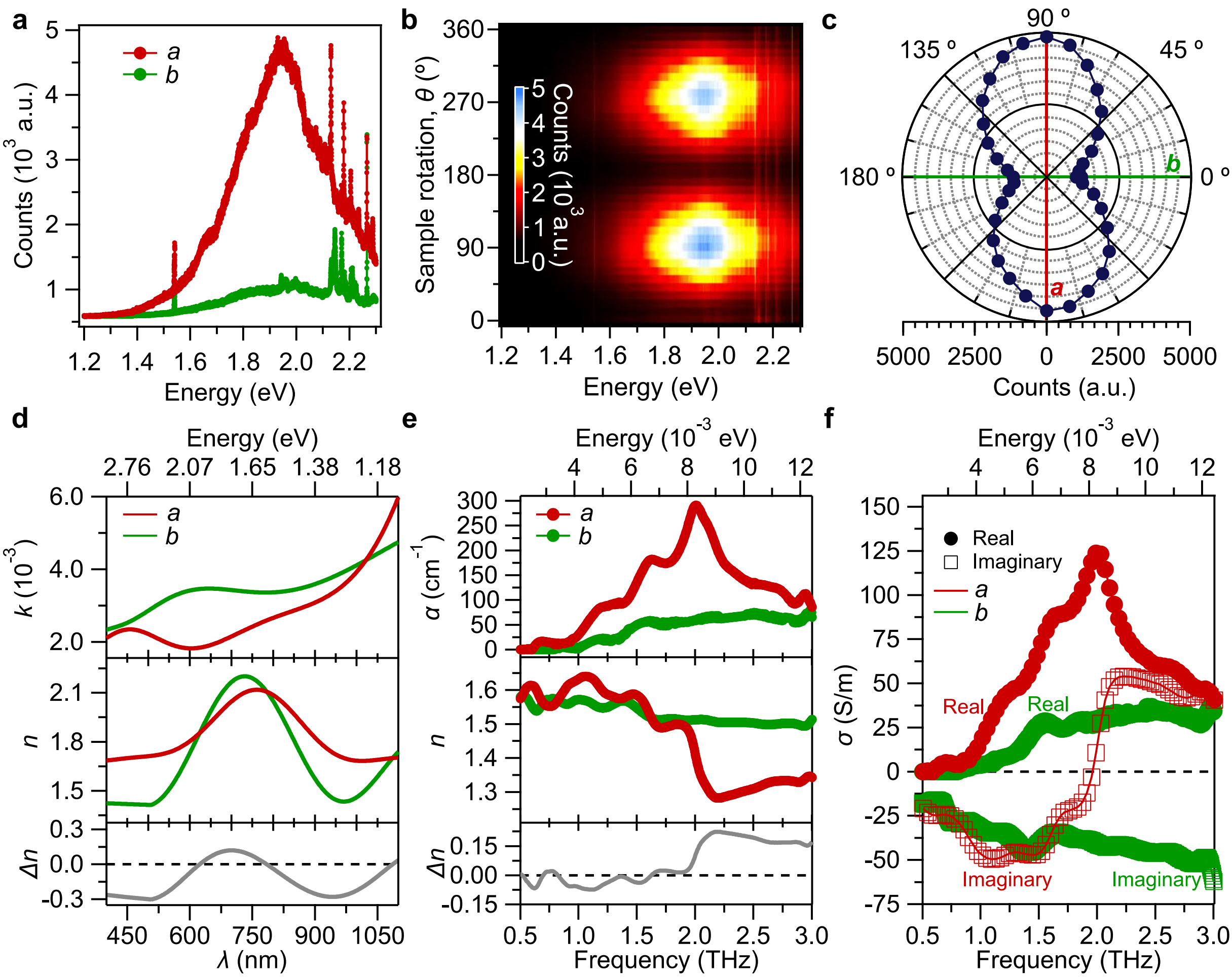

**Figure 2. Anisotropy in the optical properties of [FeCl(pzCl)(bpy)].** a-c) Photoluminescence resolved spectra with linearly polarized incident light aligned along different crystal orientations, showing the limiting cases for the *a* and *b* crystallographic axis (a), a colormap with the full angular dependence in the range of 1.2-2.3 eV (b) and a polar plot at 1.96 eV (c), where 0° (90°) correspond to the *b* (*a*) axis, respectively. d) Imaginary (top panel) and real part of the refractive index (middle panel) as well as the birefringence (bottom panel), defined as the difference between the refractive index along the *b* and *a* axis, in the visible range. e) Optical anisotropy in the THz range. Absorption (top panel) and real part of the refractive index (middle panel), as well as the birefringence (bottom panel), defined as the difference between the refractive index along the *b* and *a* axis. e) Complex AC conductivity.

### 2.3. Band structure calculations

To elucidate the pronounced anisotropy in the optical properties of [FeX(pzX)(bpy)] (X = Cl and F), and the different PL behavior measured in the two materials, we performed DFT+U calculations. **Figure 3** shows the computed band structures of the two derivatives, with the projected contributions from the 4-chloropyrazolate (green) and 4,4'-bipyridine (red) ligands. A band gap of *ca.* 0.1 eV for the Cl derivative and 0.3 eV for the F derivative is predicted. Both values are significantly smaller than the expected optical band gap based on our PL measurements. This underestimation is a well-known limitation of semi-local exchange-correlation functionals such as PBE, and is particularly pronounced in MOFs, where highly localized metallic centers interact with extended and delocalized organic ligands.[73,74] Nevertheless, calculations at this level of theory can provide valuable microscopic insight into the mechanisms governing the anisotropic electronic behavior.

The electronic structure of the Cl derivative (**Figure 3.a**) shows that the valence band maximum (VBM) is located along the Z–D path in reciprocal space and has predominantly a 4,4'-bipyridine character. The conduction band minimum (CBM), on the other hand, appears as a nearly flat band along the Γ–Z–D–B path, with the absolute minimum located at Γ. Although this would technically correspond to an indirect band gap, the energy difference between the CBM at Γ and along Z–D is minimal (of the order of a few meV), and thus within the expected uncertainty of GGA-level calculations. [73,74] Given this, and considering the known limitations of PBE in resolving subtle energetic differences, the result can still be considered compatible with the observed PL in the Cl derivative. On the other hand, in the F derivative the VBM retains a similar character, but the CBM is qualitatively different: it is formed by a more strongly hybridized band along the Z–D direction, leading to a clearly indirect band gap. These theoretical insights align with PL measurements, which indicate a transition from direct to indirect band gap upon chemical substitution of Cl by F.

As far as the electronic anisotropy is concerned, we note that both materials display the most dispersive bands along the Γ–A and $E–Z–C_2–Y_2$ paths, corresponding to directions parallel to the *a*-axis in real space. This observation is consistent with the strong electronic anisotropy measured experimentally, as discussed in the previous section, highlighting the crystal structure as the origin of this behavior. In line with this observation, we calculate the phonon dispersion for FeCl(pzCl)(bpy) (**Supplementary Section 7**), revealing a larger phonon dispersion along Γ–A and E–Z–C2–Y2 paths (*a* axis). This aligns with a pronounced structural anisotropy, since a higher phonon dispersion is related to a higher rigidity of the structure along this direction and, therefore, is compatible with the different deformation capabilities of bipyridine and pyrazolate ligands, which are aligned to *a* and *b* axis, respectively.

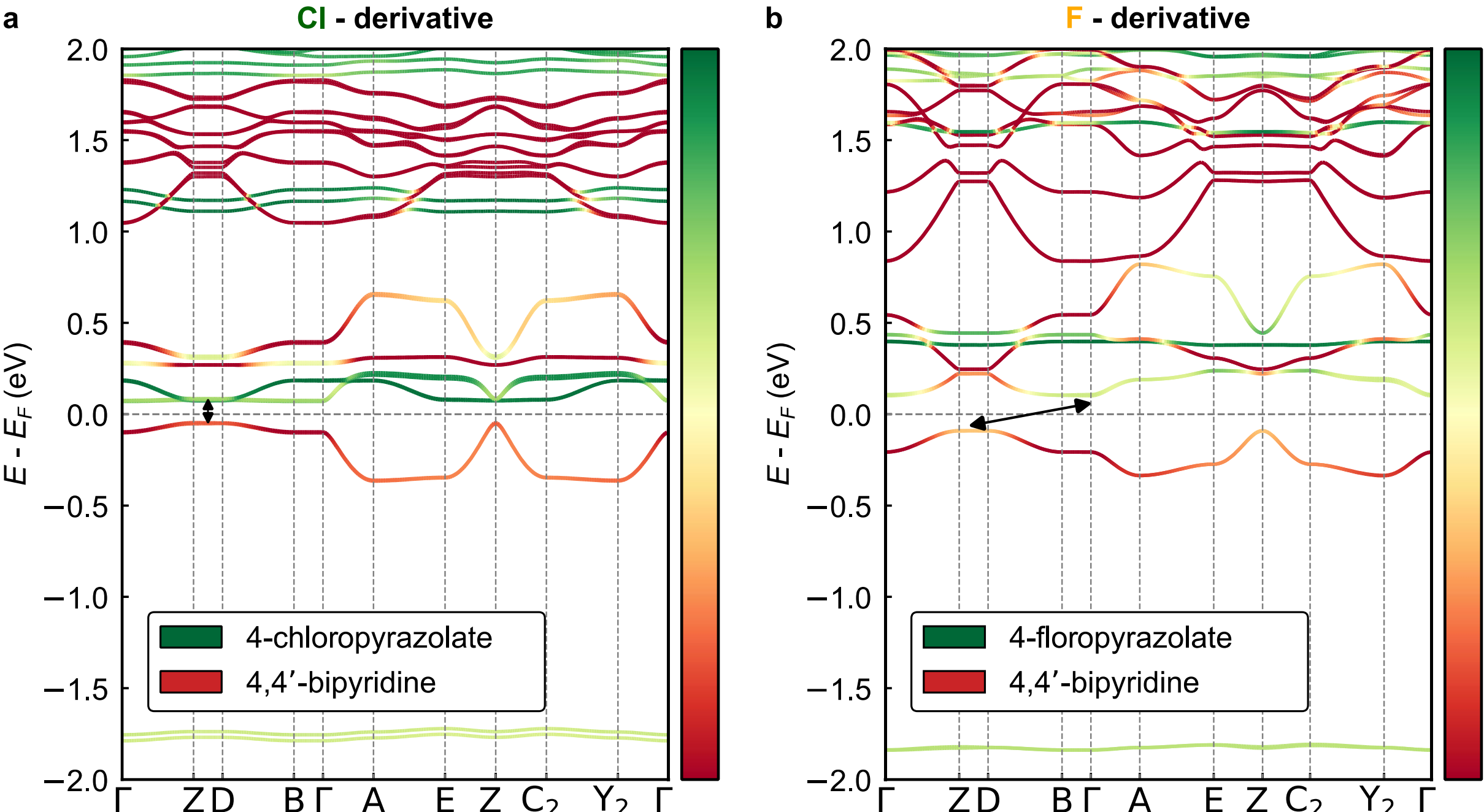

**Figure 3. Band structure calculations of [FeX(pzX)(bpy)] (X = Cl and F).** The color scale encodes the contribution of the different ligands in the structure (pzX in green and bpy in red). A black arrow highlights the difference between the flat band in the conduction band in Z-D of the Cl-derivative if compared to the F-derivative.

### 2.4. Exfoliation of [FeX(pzX)(bpy)] and tuning of the optical properties in twisted heterostructures

Atomically-thin layers of these layered MOFs have been obtained by the mechanical exfoliation of the bulk counterpart, as typically performed with graphene and other 2D materials. In **Figure 4.a**, thin-layers of the Cl derivative deposited on top of a $SiO_2$/Si substrate are shown, with thickness that range from a few to hundreds of nanometers. We present in **Figure 4.b** an atomic force microscopy image of a flake with different thicknesses and with the corresponding height profiles in **Figure 4.c**, being layers with thickness below 10 nm achievable. Interestingly, when illuminated with a linearly polarized incident light, the color (top panel in **Figure 4.d)** and optical contrast *C* (right panel in **Figure 4.d**) is highly dependent on the relative orientation between the light polarization and the crystalline axis, where 0° (left panel) and 90° (right panel) correspond to horizontally and vertically polarized light, respectively. For example, when the incident linearly polarized light is along the *b* (*a*) axis, the flakes exhibit a dark reddish (greenish) color with negative (positive) optical contrast. Considering the bare substrate, no significant dependence is observed. This effect can be attributed to multiple internal reflections (etaloning) and modelled based on the Fresnel equations (see **Supplementary Section 4.2**). Within the Fresnel law, the optical contrast depends on the sample thickness, the refractive indices of the involved media, and the illumination wavelength. For the commonly used 285 nm $SiO_2$ substrate, as employed in this work, the optical contrast change is theoretically and experimentally shown to be minimal for ultra-thin layers (as we develop in the **Supplementary Section 4.2** and experimentally observe in the **Supplementary Videos**). Nonetheless, by tuning the substrate thickness and incident wavelength (for example, using 1000 nm $SiO_2$ and illumination at 460 nm or 875 nm) ultra-thin layers can be made optically visible, as calculated in **Figure S57d**. The full angular contrast dependence is presented in **Figure 4.e**, and we show the same effect for other flakes with different thickness in the **Supplementary Video 1** and **Supplementary Section 4.1,** being the effect especially remarkable for flakes with thicknesses comparable to the incident wavelength. A similar effect is observed for the F-derivative (**Figure 4.f-g, Supplementary Video 2** and **Supplementary Section 4.3**).

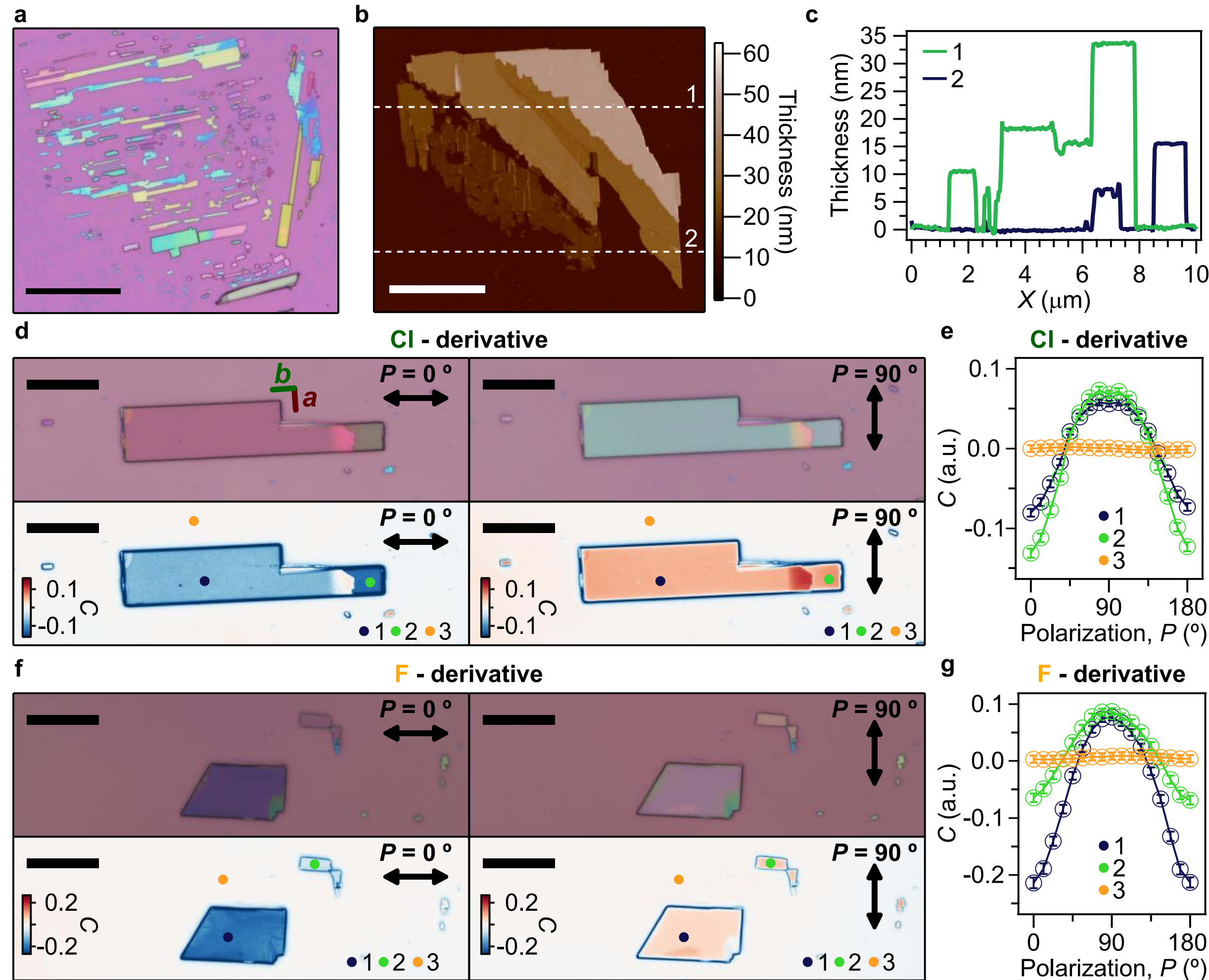

**Figure 4. An exfoliable van der Waals MOF [FeX(pzX)(bpy)] (X = Cl and F).** a) Optical image of mechanically exfoliated flakes of [FeCl(pzCl)(bpy)] on top of a 285 $SiO_2$/Si substrate. Scale bar: 60 μm. b) Atomic force microscopy image of an exfoliated flake in a. Scale bar: 3 μm. c) Height profiles at the dashed areas marked in b. d-e) Optical images (top panel) for the chlorine derivative compound with the calculated optical contrast (bottom panel) with an incident linearly polarized light at 0° (left panel) and 90° (right panel) together with the optical contrast angular-dependence upon different incident linearly polarized light (e) at the locations indicated in d. Position 3 is the silicon substrate. Scale bar: 20 μm. f -g) Optical dependence, analogous to d-e, for the fluor derivative. The full dataset of images is shown in the **Supplementary Section 4** and **Supplementary Videos 1-2**.

These layers can be easily manipulated following the deterministic transfer method developed for 2D materials (see **Experimental Section**). As an example, we have fabricated an orthogonally-twisted vdW heterostructure for the Cl derivative by placing two-flakes on top of each other with an in-plane rotation of 90 degrees (**Figure 5.a**). When illuminated with a linearly polarized incident light, the color (top panel in **Figure 5.a**) and optical contrast *C* (right panel in **Figure 5.b**) of the two flakes (denoted as 1 and 2) exhibit 90 degrees offset, as result of the orthogonal structure. Considering the twisted region (denoted as 3), there is a slight optical contrast angular dependence as well but, for all the angular range, it remains negative. The full dataset of images is shown in the **Supplementary Video 3**. The optical behavior is correlated with the spectral behavior by performing differential reflectivity measurements (see **Experimental Section**) in the pristine flake and twisted areas, as shown in **Figure 5.c**. In order to optimize the reflectivity variation, the thickness of the flakes (850 and 870 nm; **Supplementary Section 4.1.11**) has been chosen to be comparable to the wavelengths used (visible range; see **Experimental Section**). Under these conditions, the normalized reflectivity spectra in the visible range presents a characteristic chromatic dependence arising from etaloning effects, with a significant variation when the polarization is aligned with the *b* (0°) or *a* (90°) axis (top panel in **Figure 5.c**). This contrasts with the case of the twisted area, where only minor variations are observed (bottom panel in **Figure 5.c**). This effect can be better visualized in the polar plots presented in **Figure 5.d** for the flake and the twisted heterostructure for two characteristic frequencies (570 nm and 650 nm, denoted with green and red dashed lines in **Figure 5.c**). For the flake at $\lambda$ = 570 nm, the normalized reflectivity is anisotropic exhibiting a clear two-fold symmetry with a maximum value that is *ca.* 2.5 times higher than the substrate reflectivity. On the contrary, at $\lambda$ = 650 nm, the reflectivity is almost isotropic. Considering the twisted heterostructure, the large anisotropy in the reflectivity observed for the pristine flake is suppressed, suggesting that the etalon formed from the cross section of two flakes with similar thickness compensates the optical anisotropy of its constituent flakes. Finally, we consider the optical response with a cross-polarized configuration (**Figure 5.e**). Upon rotating the sample, the bare flakes exhibit an optical contrast for angles different from 0° and 90° (that is, when the analyzer/polarizer is not aligned with the crystallographic axis), with a clearly suppressed signal in the twisted area (**Figure 5.f**). The full angular dependence is shown in the **Supplementary Video 4**. Similar trends are observed for a second orthogonally-twisted vdW heterostructure based on thinner layers (thickness of 16 nm and 19 nm, **Supplementary Section 4.1.11**) but with a smaller signal (**Supplementary Section 4.1.8** and **Supplementary Video 5-6**).

Overall, we demonstrate the possibility of modifying the intrinsic optical anisotropy of the layered MOF material when combined in an orthogonally-twisted heterostructure (switching from an anisotropic response in the bare material to an isotropic one in the heterostructure), opening future possibilities of control by selecting the substrate as well as the thickness and twist-angle of the flakes, with potential in the design of integrated photonics and optoelectronic devices.[75–77]

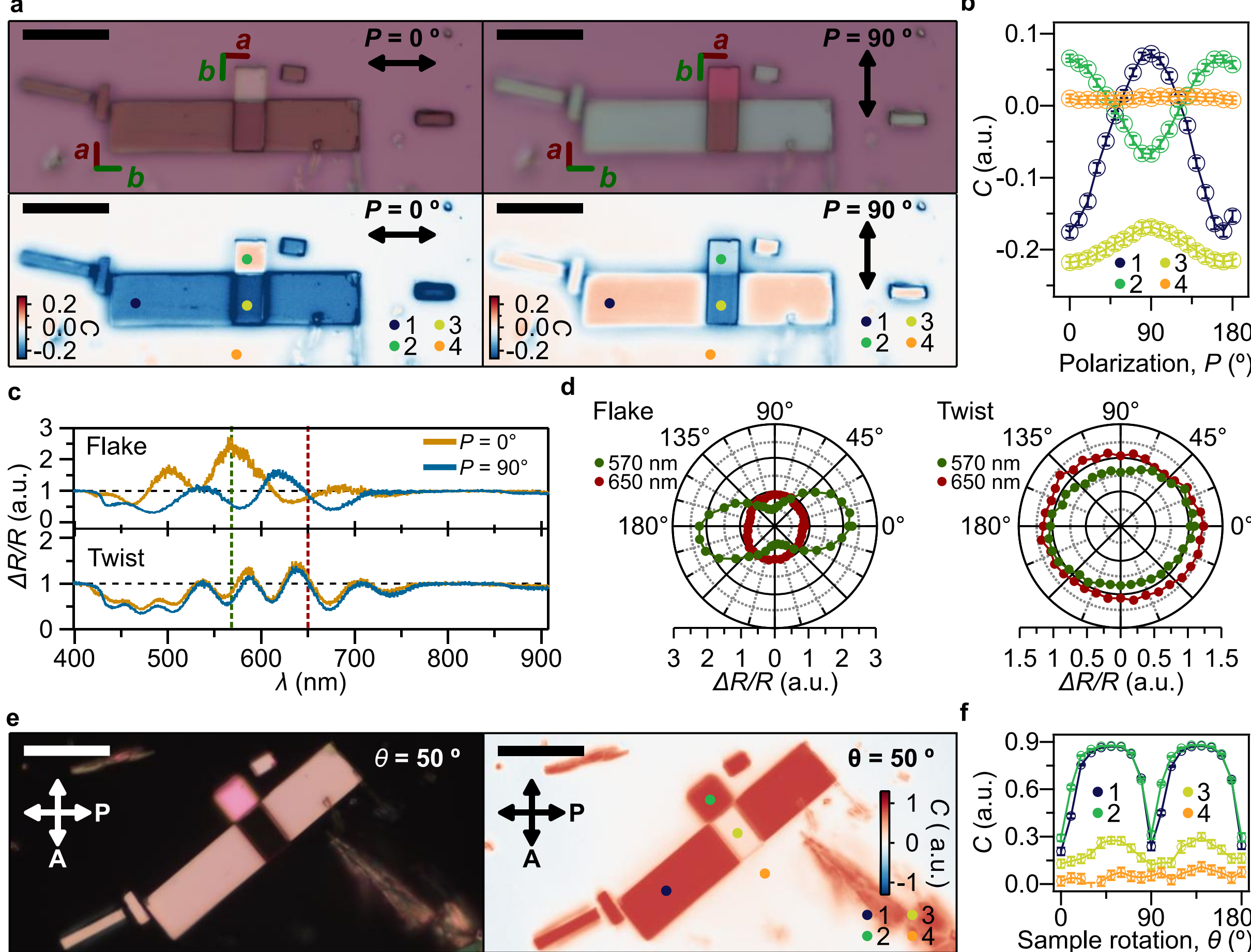

**Figure 5. A twistable van der Waals MOF [FeCl(pzCl)(bpy)].** a) Optical images (top panel) with the calculated optical contrast (bottom panel) for an orthogonally-twisted heterostructure of [FeCl(pzCl)(bpy)] with an incident linearly polarized light at 0° (left panel) and 90° (right panel). Scale bar: 20 μm. b) Optical contrast angular-dependence upon different incident linearly polarized light at the locations indicated in a. Position 4 is the $SiO_2$ substrate. The full dataset of images is shown in the **Supplementary Video 3**. c) Differential reflectivity for the flake and twisted area measured at the positions denoted as 1 and 3 in a, respectively. d) Angular dependence of the differential reflectivity upon different incident linearly polarized light for the flaked and twisted area at 570 nm and 650 nm, as highlighted with vertical dashed lines in c. e)

Optical (left panel) with the calculated optical contrast (right panel) for an orthogonally-twisted heterostructure with an incident cross-polarized light and a rotation angle of 50°. Scale bar: 20 µm. f) Optical contrast angular-dependence with a cross-polarized incident light upon different sample rotation angles at the locations indicated in e. Position 4 is the silicon substrate. The full dataset of images is shown in the **Supplementary Video 4**.

## 3. Conclusion

We have demonstrated a molecular strategy for developing 1D materials embedded in 2D MOFs, as exemplified by the [FeX(pzX)(bpy)] (X = Cl, F) family. These layered systems exhibit chain-like arrangements of iron centers along the *b*-axis, resulting in marked structural and optical anisotropy, as confirmed by optical measurements in the visible and THz ranges and supported by band structure calculations. In addition, thanks to the unconventional way for preparing these MOFs – involving a solvent-free method instead of a solvothermal method – large crystals of high quality have been obtained. This has been essential for the ulterior exfoliation and manipulation of the van der Waals layers using the common methods developed for handling 2D materials. In terms of the materials properties, the pronounced optical anisotropy is reflected in the large optical birefringence observed both in visible and THz ranges, which is comparable with other birefringent materials. Interestingly, the intrinsic optical anisotropy of these layered MOF materials can be tuned and even switched-off in the visible range by fabricating vdW heterostructures formed by orthogonally-twisted thin layers. In the context of van der Waals materials, these molecular-based systems represent one of the very few examples of mechanically exfoliable MOFs and the first attempt of modulating their properties *via* twist-engineering.

Overall, this molecular approach exploits the synergies between the chemical versatility of MOF materials, providing a structural platform to stabilize electronically-isolated molecular chains in vdW layers, and the versatility of 2D materials in terms of stack- and twist-engineering, opening a convenient way to explore, manipulate and exploit the properties of 1D materials. As compared with a purely inorganic approach, the molecular approach provides a more accessible and versatile alternative for tuning the properties of these vdW semiconductors. In perspective, this work also highlights the potential of chemical design in metal-organic van der Waals semiconductors, where the material's properties can be tuned by choosing selectively the molecular ligands, the halogens and the metals. Thus, a chemical engineering of these layered MOFs should allow us to tune the 1D character of these low dimensional materials in the future by using molecular bridges of varying lengths to link the metal chains. On the other hand, a compositional engineering will offer the opportunity to explore other functionalities, such as the magnetic ones. For example, 1D chains with controlled spin and optical anisotropy may be isolated in these 2D coordination networks by changing $Fe^{2+}$ with Co, Ni or Mn, which can be integrated with other 2D materials for fabricating vdW heterostructures of interest in spintronics and optoelectronics.

## 4. Experimental Section

*Synthesis of [FeX(pzX)(bpy)] (X = F and Cl):* Ferrocene (30 mg, 0.16 mmol), 4-halopyrazole (30 mg and 35 mg for pzF and pzCl, respectively, 0.34 mmol) and 4,4'-bipyridine (50 mg, 3.12 mmol) were combined and sealed under vacuum in a layering tube (4 mm diameter). The mixture was heated at 250 °C for 3 days to obtain air-stable, brownish, transparent, plate-like crystals suitable for X-ray single-crystal diffraction studies. The resulting products were cleaned with acetonitrile and acetone, and the main compounds were isolated on a 60 % yield. Phase purity was established by X-ray powder diffraction.

*X-ray Diffraction:* Single crystal X-ray diffraction (SCXRD) studies were performed on a Rigaku Oxford Diffraction Supernova diffractometer (Mo) X-ray source ($\lambda$ = 0.71073 Å) for [FeCl(pzCl)(bpy)] and a DW rotating anode synergy R diffractometer with (Cu-$K_{\alpha}$) X-ray source ($\lambda$ = 1. 54184 Å) for [FeF(pzF)(bpy)]. In both cases, a single crystal was mounted on glass fibers using a viscous hydrocarbon oil to coat the crystals and then transferred directly to the cold nitrogen stream for data collection. Data were measured using the CrysAlisPro suite of programs. The structure was initially solved using the SHELXT program[78] implemented through OLEX2 (v1.5)[79] and refined using SHELXL[80] least squares refinement procedures against all $F^2$ values. Solvent mask protocol implemented in OLEX2 was used to account for the residual electron density corresponding to 4,4-bipyridine molecules located within the pores in both structures. The diffuse 4,4-bipyridine molecules could not be assigned and were not included in the formula. Consequently, the molecular weight and density given above are underestimate. Crystallographic information files CCDC 2482068 and 2497274 contain full details for all crystal structures reported. For Powder X-Ray Diffraction (PXRD) experiments, the samples were lightly ground in an agate mortar and pestle and used to fill 0.7 mm borosilicate capillaries that were mounted and aligned. For [FeCl(pzCl)(bpy)] sample, an X-ray diffractometer (PANalytical Empyrean) with copper as a radiation source (Cu $K_\alpha$, 1.5418 Å) was used. [FeF(pzF)(bpy)] samples were measured on a Bruker D8 Discover powder diffractometer, using Cu $K_\alpha$ radiation ($\lambda$ = 1.54056 Å). The detector was an EIGER2 R 500K, multi-mode 2D

*Differential reflectivity measurements:* The optical characterization of microcrystals is carried out using a regular optical microscope provided with an additional collection path for spectroscopy. The same tungsten lamp used for illumination proposes is used as white light source in the differential reflectivity measurements. A polarization analyzing system composed of a half-wavelength plate and a Polarizer is located the optical collection path. Afterwards, a non-polarizing beam splitter is used to deviate the light reflected on the sample surface by two different channels: the first one used for imaging and the and the second one used for to focus the collection on a multimode optical fiber (50um core) connected to a spectrometer. In the case of reflectivity measurements, the operation is limited to the visible range due to spectra of the illumination lamp together with the antireflective coatings of the optical components limits and the reflectivity of the silica on silicon substrate which is used as a reference.

*Transmissivity of bulk crystals:* The transmissivity of bulk crystals is carried out using an optical ellipsometer from SOPRA. This system allows the use of several photometry techniques in the rage from 230-1100 nm wavelength. It is provided of a microspot that allows measuring transmissivity on crystals with lateral size of few milimiters.

*Terahertz spectroscopy:* THz time domain spectroscopy (THz-TDS) was performed by a homemade setup feed with a Ti:Shaphire laser (800nm, 50fs, 1KHz). THz generation was accomplished by optical rectification on a 1mm ZnTe crystal. The detection of the THz freely propagating and transmitted pulses was realized by electro-optic sampling, also employing a 1 mm ZnTe crystal and a Si based balanced photodiodes. An optical profilometer Profilm3D was employed to measure the sample's thicknesses.

*Computational details:* First-principles DFT calculations were performed using SIESTA code. We used GGA+PBE method to describe the exchange-correlation energy. Hubbard U corrections ($U_{eff} = 6$ eV) as implemented in SIESTA were considered for the strongly correlated Fe 3d electrons. We used norm-conserving scalar relativistic pseudo-potentials taken from the Pseudo-Dojo database in the psml format. Grimme D3 dispersion corrections were applied to consider for vdW interactions. A real-space mesh cutoff of 900 Ry and a 3 × 5 × 3 Monkhorst–Pack k-point mesh was used in all calculations, in combination with double-ζ basis set for all atoms. All structures were relaxed until the forces were less than 0.04 eV $Å^{-1}$ in all atomic coordinates.

*Exfoliation, van der Waals heterostructure fabrication and characterization:* 2D layers were mechanically exfoliated using adhesive tape (80 μm thick adhesive plastic film from Ultron Systems) and placed on top of 285 nm $SiO_2$/Si substrates (from NOVA Electronic Materials, LLC). The exfoliated samples were inspected by optical microscopy (Nikon Eclipse LV-100 optical microscope with a Nikon TU Plan Fluor 100× objective lens of 1 mm working distance with a linear polarizer), atomic force microscopy (Nanoscope IVa from Veeco) and Raman and photoluminescence measurements (LabRam HR Evolution confocal Raman microscope from Horiba using 100× objectives Olympus LMPlanFL N and a grating of 600 gr $mm^{-1}$ upon different incident wavelengths as shown in the **Supplementary Section 2.1**). The van der Waals heterostructures were built in a deterministic way using polycarbonate and the help of a micromanipulator. The optical contrast is calculated from optical microscopy images and defined as $C = (I_{flake} - I_{substrate})/( I_{flake} + I_{substrate})$, where the $I_{flake}$ value is obtained by selecting a region of interest in an RGB image and averaging its intensity. The same procedure holds for $I_{substrate}$.

**Acknowledgements**

The authors acknowledge financial support from the European Commission (PATHFINDER 4D-NMR 101099676 and NextGenerationEU/PRTR, ERC-2021-StG 101042680 2D-SMARTiES), the Spanish MCIN (Projects 2D-SPICE PID2023-149309OB-I00 and 2DM PID2022-137078NB-I00, cofinanced by FEDER; Excellence Unit “María de Maeztu” CEX2024-001467-M and “Severo Ochoa” CEX2020-001039S Programmes for Spanish Centers and Units of Excellence in R&D; and the Consolidación Investigadora 2022 program CNS2022-136203), the Generalitat Valenciana (PROMETEO Program, PO FEDER Program IDIFEDER/2021/078, CIDEGENT/2018/005, CIDEXG/2023/1). This study forms part of the Advanced Materials and Quantum Communication programmes and was supported by MCIN with funding from European Union NextGenerationEU (PRTR-C17.I1) and by Generalitat Valenciana (projects MFA/2022/050, COMCUANTICA/010, and COMCUANTICA/011). V.B. acknowledges financial support from the Comunidad de Madrid under the project "Cesar Nombela"2023-T1/TEC-29119. M.B thanks the funding received from the grant JDC2022-048249-I funded by MICIU and by the “European Union NextGenerationEU/PRTR”.

## Supporting Information

Supporting Information is available from the Wiley Online Library.